\documentstyle[12pt,a4]{article}
\input{epsf}

\newcommand{\R}{I \! \! R}

\newcommand{\beq}[1]{\begin{equation}\label{#1}}
\newcommand{\eeq}{\end{equation}}

\begin{document}
\begin{titlepage}
\hfill{Freiburg THEP-97/18}
\bigskip
\bigskip

\begin{center}
{\huge Quantization of Black Holes in the Wheeler-DeWitt Approach}
\bigskip
\bigskip
\bigskip

Thorsten Brotz\footnote{e-mail: brotz@physik.uni-freiburg.de}
\bigskip
\bigskip

Fakult\"at f\"ur Physik der Universit\"at Freiburg, Hermann-Herder-Stra{\ss}e
3, \newline
D-79104 Freiburg, Germany
\bigskip
\bigskip
\bigskip

\begin{abstract} We discuss black hole quantization in the Wheeler-DeWitt
approach. Our consideration is based on a detailed 
investigation of the canonical formulation of gravity with special 
considerations of surface terms. Since the phase space of gravity
for non-compact spacetimes or spacetimes with boundaries is ill-defined
unless one takes boundary degrees of freedom into account, we give a Hamiltonian 
formulation of the Einstein-Hilbert-action as well as a Hamiltonian 
formulation of the surface terms. 
It then is shown how application to black hole
spacetimes connects the boundary degrees of freedom with thermodynamical
properties of black hole physics. Our treatment of the surface terms
thereby naturally leads to the Nernst theorem. Moreover, it will produce
insights into correlations between the  Lorentzian and the Euclidean
theory. Next we discuss quantization, which we perform
in a standard manner. It is shown how the thermodynamical properties can 
be rediscovered from the quantum equations by a WKB like approximation 
scheme. Back reaction is treated by going beyond the first order 
approximation. We end our discussion by a rigorous investigation of
the so-called BTZ solution in $2+1$ dimensional gravity. 
\end{abstract}
\bigskip
\bigskip
\bigskip

Submitted to Physical Review D
\end{center}
\end{titlepage}
\section{Introduction}

Most papers about canonical quantum gravity are dealing with spatially
compact systems without boundaries. This allows one to skip boundary terms
that would otherwise
arise from the variation of the Einstein-Hilbert action. On the other hand,
the importance of these surface terms was already emphasized by DeWitt
\cite{DeWitt} and
exhaustively discussed for spatially asymptotic spacetimes in
\cite{ReTe, BeNo}.
If the spacetime is, for example, not compact, one must take care of the
definition of phase space.
In general relativity a correct definition can be achieved only if one adds 
appropriate boundary integrals to the Einstein-Hilbert action.
A correct definition of the classical phase space is, of course, a crucial
step for the canonical quantization procedure. Thus, a canonical formulation
of quantum gravity for spacetimes with boundaries or non-compact spatial
regions, e.g. black holes,
must also include a discussion of the corresponding surface terms.
In the spherically symmetric sector of general relativity this was 
recently illustrated by Kucha\v{r} \cite{Kuchar}. In \cite{Kuchar} Kucha\v{r} explicitly shows, that the 
reduction of the parametrized action for primordial Schwarzschild black 
holes to true dynamical degrees of freedom is only achieved if boundary 
terms (in this case the ADM-mass at infinity) are taken into account.

As is well known, surface terms play also a key role in black hole 
thermodynamics. This was first discussed by Gibbons and Hawking
\cite{GiHa} who could derive the thermodynamical properties of black holes
by a path integral formulation of quantum gravity. They found that in the
saddle point approximation the thermodynamical laws of black holes are
related to surface terms of the Einstein-Hilbert action.
A variety of investigations, exploring this relation from distinct
viewpoints, led to a deeper understanding of black hole thermodynamics
(see for example \cite{CaTe, Teitelboim, BrKi, Brownb, BoPaPe, SuWa} and 
references therein).
Moreover, a statistical interpretation can be derived
in 2+1 dimensional gravity \cite{Carlipb}.

In recent years  another kind of
(semiclassical) approximation scheme has been developed, which should
also offer an alternative access to a thermodynamical description of
black hole mechanics. This scheme is
formulated in the Wheeler-DeWitt approach to quantum gravity and
is strongly related to a Born-Oppenheimer kind of approximation.
It is formally given by a WKB-like
expansion of the state functional $\Psi$ \cite{Kiefer}. One big advantage
of this WKB approximation is that it connects the `ill-defined'
quantum equations
(Wheeler-DeWitt-equation) with the much better understood quantum field
theory in curved spacetimes. Moreover, it provides the theory of quantum
gravity with a semiclassical time function called WKB-time.
This WKB-time is tied to the
recovery of a functional Schr\"odinger equation for matter fields on
a curved background from the Wheeler-DeWitt equation.
Thus, {\it time} has to be viewed in this approach as a semiclassical
property. Consequently, the theory of quantum gravity must be interpreted
as `timeless' on the full level.
Until now only little attention has been paid to boundary terms in
formulating the WKB approximation. On the other hand, it was emphasised
recently that for the recovery of black hole thermodynamics
from the WKB approximation boundary terms have to be included \cite{BrKi}.

In this paper we shall give a constrained formulation of the `boundary
states' and illustrate how this leads to a consistent canonical
quantum formulation. We will concentrate on the case of `black hole
boundaries', since we are mainly interested in the discussion of
thermodynamical properties. Extension to other boundaries can easily
be achieved. The study will generalize the results of
\cite{BrKi}, where only spherically symmetric black holes were
discussed. Our classical description will uncover relations of 
the boost parameter at the bifurcation point and
the opening angle of the Euclidean spacetime with the surface gravity
of black holes. Moreover, the Nernst theorem for black hole mechanics
will appear as a simple consequence of the obtained relations.
For quantization the used constrained 
formulation of the boundary terms will prove to be very advantageous.
Nevertheless, interpretational problems still appear. These problems are
avoided in the above mentioned WKB approximation. 
The classical thermodynamical description can be found in the highest order
of this approximation scheme. Moreover,
we will demonstrate how back reaction influences the entropy of
black holes.

Our paper is organized as follows.
In section 2 we give a brief review of the Hamiltonian formulation of the 
Einstein-Hilbert action for compact spacetimes $\cal M$ with topology $\Sigma
\times \R$ and boundary $\partial {\cal M}$ following, with elaborations, 
the work of \cite{HuHa}. Beyond that we explicitly discuss the
Hamiltonian formulation of the
surface terms, since this is crucial for our further investigation.
In section 3 the application of section 2
for the particular case of black hole spacetimes is discussed. We shall
demand $\cal M$ to lie within the right wedge of the corresponding Kruskal
diagram, and its boundary $\partial {\cal M}$ to include the bifurcation
two-sphere as well as spatial infinity, where we assume our theory to
be Poincar\'e invariant. Section 4 is devoted to thermodynamical
considerations that will relate our Hamiltonian treatment of the surface terms
with the standard results found by other methods. It includes an explicit 
computation of the boundary term
at the bifurcation point. This computation will connect the boost parameter
at the bifurcation point \cite{Hayward} with the surface gravity of
the black hole. Next we show  the relation between the boundary degrees
of freedom and the first law of black hole mechanics by using perturbational
methods discussed in \cite{Wald}. Moreover, we give a proof of the Nernst theorem for black holes, which was for a long time
thought to be violated \cite{WaldRe}. Since a similar proof of the
Nernst theorem was found in the Euclideanized theory of gravity
recently \cite{Teitelboim,GiKa,HaHo}, we will relate in section 5 our results
with those of the Euclidean approach.  We also comment on topological considerations
in the realm of the Lorentzian theory which were recently discussed to
explain the meaning of entropy for gravitating systems 
\cite{BrKi,Martinez}.
In section 6 we pass over to the quantum theory by Dirac's constrained
quantization method. In contrast to the standard midisuperspace 
quantization this will now inculde also boundary constraints.\footnote{For 
a review of the state of art in the midisuperspace quantization of black 
holes see \cite{Kuchar}.} 
Since it is still unclear how to extract thermodynamical properties
from the full quantum equations we shall overcome this problem in section 7 by
giving an appropriate semiclassical approximation. This semiclassical
approximation scheme connects the WKB-phase of the wavefunction with
the classical laws of black hole mechanics.
Going beyond the first order of approximation leads to back reaction
corrections. 
In section 8 we show how our treatment works in the particular example of
$2+1$ gravity. We start with a purely classical description of the 
so called BTZ black hole. This will include a derivation of the 
Hamilton-Jacobi functional. Then the case of conformally 
coupled matter fields on the background of the BTZ solution is discussed.
Moreover, it is shown how back reaction leads to corrections of 
entropy and temperature. The derived entropy, which follows directly
from the deduced boundary terms, differs from the one given in
\cite{MaZa}. The reason for this difference is found to follow from
contributions of the conformal coupling (which are included in our
treatment, but not in \cite{MaZa}). These contributions are important, 
since otherwise the first
thermodynamical law for black holes $dM = T dS$ does not hold.  
Finally we shall briefly comment on results in string theory,
where recently a statistical interpretation of black hole thermodynamics 
was given (for a review see e.g. \cite{Horowitz}), and where a different 
conclusion has been drawn 
about the behaviour of the entropy in the extremal cases.

\section{Action and ADM Decomposition}

Let ${\cal M}$ denote a four manifold of topology $\Sigma \times \R $
endowed with a Lorentzian metric $g_{\mu\nu}$. The derivative operator
associated with $g_{\mu\nu}$ is called $\nabla_\mu$. The boundary of ${\cal M}$
consists of initial and final spacelike hypersurfaces $\Sigma_0$ and
$\Sigma_1$, and a timelike hypersurface $B=B_t \times \R$ [Fig. 1].
The induced metric on the spacelike hypersurface $\Sigma$ is denoted by
$h_{ab}$, the induced metric on the timelike boundary $B$ by
$\gamma_{ij}$, and the induced metric on the edges $B_0$ and $B_1$
by $\sigma_{ab}$.
The Einstein-Hilbert action of general relativity is in
this case given by \cite{Hayward}:
\begin{eqnarray}  \label{1}
I[{\cal M}, g] &=& \frac{1}{16\pi G} \int_{\cal M} d^4x \sqrt{-g} \left(
{\cal  R} - 2\Lambda \right) + \frac{1}{8\pi G} \int_{\Sigma_0}^{\Sigma_1} d^3x
\sqrt{h} K \\
& & + \frac{1}{8\pi G} \int_B d^3x \sqrt{-\gamma} \Theta
    + \frac{1}{8\pi G} \int_{B_0}^{B_1} d^2x \sqrt{\sigma}  \sinh^{-1}\eta
    \nonumber,
\end{eqnarray}
where $\Lambda$ is the cosmological constant.
The integral between $\Sigma_0 (B_0)$ and $\Sigma_1 (B_1)$ is an
abbreviation for the integral over the hypersurface  $\Sigma_1 (B_1)$
minus the integral over the hypersurface $\Sigma_0 (B_0)$. $K$ denotes the
trace of the extrinsic curvature $K_{ab}$ of the boundaries $\Sigma_0$
and $\Sigma_1$, and $\Theta$ the trace of the extrinsic curvature
$\Theta_{ij}$ for the boundary $B$. The final term of our action (\ref{1})
has to be included as a joint correction term since we are dealing with
spacetimes with non-smooth boundaries denoted by $B_0$ and $B_1$
\cite{Hayward}. Thereby
\beq{2}
\eta=n_\mu u^\mu
\eeq
measures the non-orthogonality of the unit normals $n^\mu$ to $\Sigma_t$
and $u^\mu$ to $B$. The ambiguity in the orientation of $n^\mu$ and $u^\mu$
is fixed by requiring that they will be future pointing and outward
pointing, respectively.

The action (\ref{1}) is only well defined for spatially compact geometries
but is divergent for noncompact ones. In the latter case one first has to 
fix a reference background $({\cal M},g_0)$ and then to consider the 
action
$I_{reg}[{\cal M}, g]=I[{\cal M},g] - I[{\cal M},g_0]$ \cite{HaHo}.
\begin{figure}[t]
\mbox{
\epsfxsize=7.0cm
\epsfbox{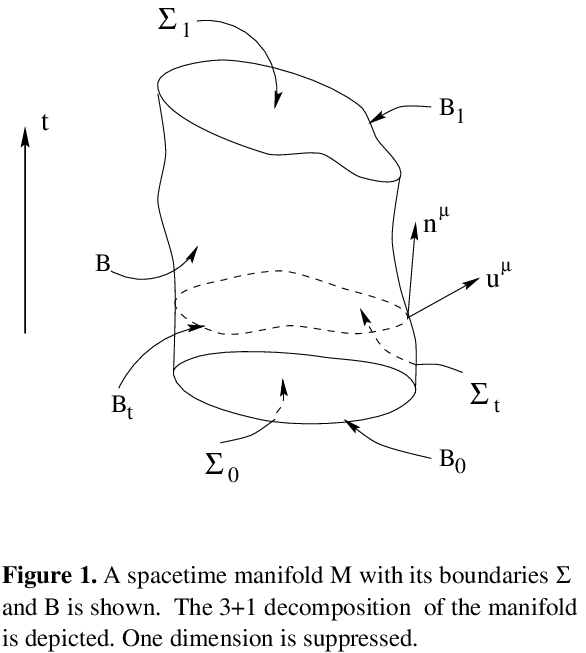}
\hspace{0.3cm}
\epsfxsize=6.7cm
\epsfbox{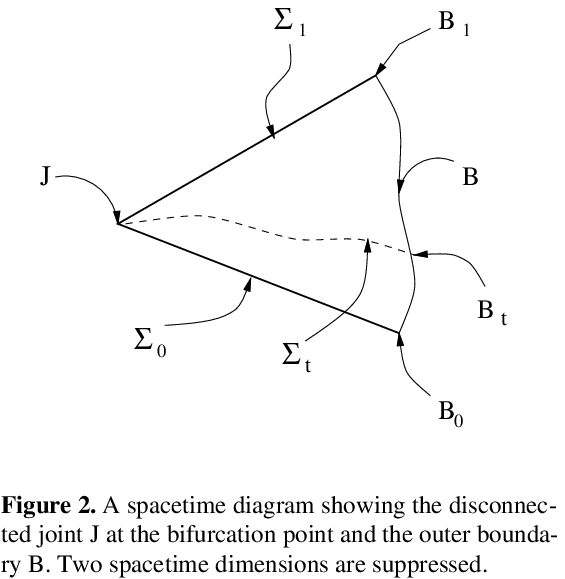}}
\end{figure}

For the ADM decomposition we write the line element of ${\cal M}$
in the standard form
\beq{3}
ds^2=-(N^2-h_{ab}N^aN^b) dt^2 + 2 h_{ab} N^a dx^b dt + h_{ab} dx^a dx^b,
\eeq
where $N$ and $N^a$ denote as usual the lapse function and the shift
vector. The scalar curvature ${\cal R}$ can be rewritten by use of the
Gauss-Codazzi relations  as
\beq{4}
{\cal R} = R + K_{ab}K^{ab} - K^2 + 2\nabla_\mu(n^\mu K - a^\mu),
\eeq
where $a^\mu=n^\nu \nabla_\nu n^\mu$ is the acceleration of the unit normal
$n^\mu$, and $R$ is the scalar curvature belonging to the hypersurface
$\Sigma_t$ with induced metric $h_{ab}$. After inserting relation (\ref{4})
into the action (\ref{1}) we can convert the integral over the total
divergence in (\ref{4}) into a surface integral over the boundary of
${\cal M}$. Thus, the action reads
\begin{eqnarray}  \label{5}
I[{\cal M}, g] &=& \frac{1}{16\pi G} \int_{\cal M} d^4x \sqrt{-g} \left[
R +K_{ab}K^{ab} - K^2 - 2\Lambda \right] \\
& & + \frac{1}{8\pi G} \int_B d^3x \sqrt{-\gamma} \left[ \Theta
    + u_\mu(n^\mu K-a^\mu) \right] \nonumber \\
& & + \frac{1}{8\pi G} \int_{B_0}^{B_1} d^2x \sqrt{\sigma}  \sinh^{-1}\eta
    \nonumber.
\end{eqnarray}
Legendre transformation then leads to the standard Hamiltonian formulation
of the action
\begin{eqnarray}  \label{6}
I[{\cal M}, g] &=& \int dt \int_{\Sigma_t} d^3x \left[p^{ab} \dot{h}_{ab}
-N {\cal H} - N_a {\cal H}^a - 2 D_a(p^{ab} N_b) \right ] \\
& & + \frac{1}{8\pi G} \int_B d^3x \sqrt{-\gamma} \left[ \Theta
    + u_\mu(n^\mu K-a^\mu) \right] \nonumber \\
& & + \frac{1}{8\pi G} \int_{B_0}^{B_1} d^2x \sqrt{\sigma}  \sinh^{-1}\eta
    \nonumber,
\end{eqnarray}
with the Hamilton constraint
\beq{7}
{\cal H} \equiv \frac{8\pi G}{\sqrt{h}} \left( 2 p_{ab}p^{ab} -p^2 \right)
          -\frac{\sqrt{h}}{16\pi G} \left( R -2\Lambda \right) \approx 0,
\eeq
and with the diffeomorphism constraints
\beq{8}
{\cal H}^a \equiv -2 D_b p^{ab} \approx 0,
\eeq
where $p^{ab} = \sqrt{h}(K^{ab}-K h^{ab})/16\pi G$ are the canonical
momenta conjugate to $h_{ab}$, and $D_a$ denotes the derivative operator
associated with the induced metric $h_{ab}$. Note that the action
(\ref{6}) is still not written in canonical form, since the boundary terms
do not possess a canonical form. Following the computation in
\cite{HuHa} we can write
\begin{eqnarray}  \label{9}
I[{\cal M}, g] &=& \int dt \int_{\Sigma_t} d^3x \left[p^{ab} \dot{h}_{ab}
-N {\cal H} - N_a {\cal H}^a \right ] \\
& & + \frac{1}{8\pi G} \int dt \int_{B_t} d^2x N\sqrt{\sigma} \left[k
    + \omega \sinh^{-1}\eta \; (\nabla_\mu v^\mu) \right] \nonumber \\
& & -2 \int dt \int_{B_t} d^2x \; r_a p^{ab}_\sigma N_b
    \nonumber,
\end{eqnarray}
with $\omega=(1+\eta^2)^{-1/2}$, with the projections $r_\mu =
\omega(u_\mu + \eta n_\mu)$ of $u_\mu$ onto $\Sigma_t$ and
$v_\mu=\omega(n_\mu-\eta u_\mu)$ of $n_\mu$ onto $B$, with $k$ denoting
the trace of the extrinsic curvature for the boundary $B_t$, and with the canonical
momenta $p^{ab}_\sigma = \sqrt{\sigma}(K^{ab}-K h^{ab})/16\pi G$ evaluated on
the boundary $B_t$.

Note that the variation of the lapse function and shift vector on the
boundary leads to the unwanted conclusion
\beq{10}
\sqrt{\sigma}\left[ k + \omega \sinh^{-1}\eta \; (\nabla_\mu v^\mu)\right]
=0,
\eeq
\beq{11}
r_a p^{ab}_{\sigma} = 0.
\eeq
This inconsistency forces one to demand that the values of the lapse function
and shift vector cannot be varied at the boundary. This corresponds to
a gauge fixation at the boundary. To get rid of this
unsatisfying gauge fixation one can perform a parametrization of 
lapse and shift \cite{Kuchar}
\beq{12}
N|_{B_t}(t) = \dot{\tau}(t),
\eeq
\beq{13}
N_a|_{B_t}(t)= \dot{\tau}_a(t),
\eeq
and treat in the following $\tau$ and $\tau_a$ as additional `boundary' 
variables.
Variation of these variables gives then the conservation laws
\beq{14}
\frac{d}{dt}
\sqrt{\sigma}\left[ k + \omega \sinh^{-1}\eta \; (\nabla_\mu v^\mu)\right]
=0,
\eeq
\beq{15}
\frac{d}{dt} r_a p^{ab}_{\sigma} = 0.
\eeq
We may now view (\ref{9}) as being written in a
mixed Hamiltonian and Lagrangian form.
In order to find the Hamiltonian formulation also for the boundary terms
one has to define the canonical conjugate momenta $\pi$ and $\pi^a$ by
a standard Legendre transformation. But this can only be consistently done
if new constraints are introduced:
\beq{16}
{\cal C} \equiv \pi - \frac{1}{8\pi G}\sqrt{\sigma}\left[ k + \omega \sinh^{-1}\eta \;
(\nabla_\mu v^\mu)\right] \approx 0,
\eeq
\beq{17}
{\cal C}^a \equiv \pi^a + 2 r_b p^{ab}_{\sigma} \approx 0.
\eeq
These constraints must be adjoined to the action by Lagrange multipliers
$N|_{B_t}$ and $N_a|_{B_t}$:
\begin{eqnarray}  \label{18}
I[{\cal M}, g] &=& \int dt \int_{\Sigma_t} d^3x \left[p^{ab} \dot{h}_{ab}
-N {\cal H} - N_a {\cal H}^a \right ] \\
& & +  \int dt \int_{B_t} d^2x  \left[\pi\dot{\tau} + \pi^a \dot{\tau}_a
    - N {\cal C} - N_a {\cal C}^a \right]. \nonumber
\end{eqnarray}
For simplicity we will introduce the notion of `outer constraint' for
constraints existing only on the boundaries and `inner constraint' for
all the others. Moreover, we shall denote the inner constraints by $\cal H$,
while for the outer constraints we shall write $\cal C$.
 
In the noncompact case one has also to take into account a background action (as mentioned
above). In this case the Hamiltonian is given by the difference between the
above computed Hamiltonian and the one computed for the background.

\section{Application to Black Holes}

In the last section we have succeeded in formulating a canonical theory that 
also takes into account additional degrees of freedom related to 
non-vanishing boundary
terms of the Einstein-Hilbert action. But we are still not at the point 
where we can treat the particular case of black holes. Until now
we have only examined joints which are embedded in three-dimensional surfaces,
whereas for black holes we have to consider isolated joints. These isolated
joints are related to hypersurfaces beginning at the bifurcation point.
Therefore, they are disconnected from the boundary belonging to the
spatially asymptotic flat region [Fig. 2].
The Einstein-Hilbert action of general relativity is in
this case given by \cite{Hayward}:
\begin{eqnarray}  \label{19}
I[{\cal M}, g] &=& \frac{1}{16\pi G} \int_{\cal M} d^4x \sqrt{-g} \left(
{\cal  R} - 2\Lambda \right) + \frac{1}{8\pi G} \int_{\Sigma_0}^{\Sigma_1} d^3x
\sqrt{h} K \\
& & + \frac{1}{8\pi G} \int_B d^3x \sqrt{-\gamma} \Theta
    + \frac{1}{8\pi G} \int_{B_0}^{B_1} d^2x \sqrt{\sigma}  \sinh^{-1}\eta
    \nonumber\\
& & + \frac{1}{8\pi G} \int_{J} d^2x \sqrt{\sigma}  \sinh^{-1}\eta
    \nonumber,
\end{eqnarray}
where the last term  takes into account the isolated joint.
Performing now the ADM decomposition as demonstrated in the last section,
the action (\ref{19}) transforms into the expression
\begin{eqnarray}  \label{20}
I[{\cal M}, g] &=& \int dt \int_{\Sigma_t} d^3x \left[p^{ab} \dot{h}_{ab}
-N {\cal H} - N_a {\cal H}^a  \right] \\
& & +  \int dt \int_{B_t} d^2x  \left[\pi\dot{\tau} + \pi^a \dot{\tau}_a
    - N {\cal C} - N_a {\cal C}^a \right] \nonumber \\
& & + \frac{1}{8\pi G} \int_{J} d^2x \sqrt{\sigma} \sinh^{-1}  \eta
    - 2 \int dt \int_{J} d^2x \; r_n ( p^{mn}_\sigma N_m) \nonumber.
\end{eqnarray}
Note that in order to fix the hypersurfaces $\Sigma_t$ at the bifurcation
point we have to demand
\beq{21}
N|_J = 0 , \hspace{1cm} N_r|_J=0.
\eeq
The vanishing of the lapse function is thereby responsible
for the cancellation of the surface terms belonging to the total derivative
in (\ref{4}) evaluated for $\Sigma_0$ and $\Sigma_1$ at the joint $J$.
On the other hand, not all diffeomorphisms are forbidden at the bifurcation point
(only those which do not generate diffeomorphisms of the bifurcation
two-sphere) \cite{Teitelboim}. The presence of these non-vanishing
diffeomorphisms explains the appeareance of the last term in (\ref{20}).
To transcribe also the boundary terms  at $J$ into canonical form we write
\beq{22}
\int_J d^2x \; \sqrt{\sigma} \sinh^{-1} \eta
 =  A(r_+) \int_{t_0}^{t_1} dt \; \frac{d}{dt} \sinh^{-1} \eta(t)
 =  \int dt\;  A(r_+) \dot{\theta}
\eeq
with the horizon area $A(r_+)=\int_J d^2x \; \sqrt{\sigma}$ and boost 
parameter $\theta = \sinh^{-1} \eta$.\footnote{In Euclidean spacetimes this
boost corresponds to a rotation and is thus also called deficit or 
opening angle.} Note that $\eta(t_0)=0$ and $\eta(t_1)=\eta$, 
since $\eta(t)$ describes the angle between the hypersurfaces $\Sigma_0$
and $\Sigma_1$. For simplicity we write
\beq{23}
\pi_J^m= - 2 r_n p^{mn}_\sigma|_J.
\eeq
Performing now the described parametrization at the boundary $J$ we end
with the canonical action
\begin{eqnarray}\label{24}
I[{\cal M}, g] &=& \int dt \int_{\Sigma_t} d^3x \left[p^{ab} \dot{h}_{ab}
-N {\cal H} - N_a {\cal H}^a  \right] \\
& & +  \int dt \int_{B_t} d^2x  \left[\pi\dot{\tau} + \pi^a \dot{\tau}_a
    - N {\cal C} - N_a {\cal C}^a \right] \nonumber \\
& & +  \int dt \left[ \pi_\theta  \dot{\theta} - \kappa {\cal C}_J
    +\int_J d^2x \left(\pi_J^m\dot{\theta}_m  - N_m {\cal C}^m_J\right)
     \right] \nonumber
\end{eqnarray}
with additional constraints
\beq{25}
{\cal C}_J \equiv \pi_\theta - \frac{A(r_+)}{8\pi G} \approx 0,
\eeq
\beq{26}
{\cal C}^m_J \equiv  \pi_J^m + 2  r_n p^{mn}_\sigma \approx 0.
\eeq
Note that $\kappa \equiv \dot{\theta} = \frac{d}{dt} \sinh^{-1} \eta$ is not
connected with the lapse function $N|_J$. Since we
are dealing with black holes, we assume covariance under Poincar\'e
transformations at infinity. Such a Hamiltonian formulation was presented
by Regge and Teitelboim \cite{ReTe}. Thus we will take over their description
and transform it to the above described constrained formulation.
As in their derivation we write for
the lapse function and the shift vector at infinity
\beq{27}
N = \alpha^0+{\beta^0}_l x^l,
\eeq
\beq{28}
N^a=\alpha^a+{\beta^a}_l x^l,
\eeq
where $\{ x^l \}$ are asymptotically cartesian coordinates and $\beta_{ab}=-\beta_{ba}$.
This leads to a split in the surface term
\beq{29}
N_a \pi^a  = -2\int_{B_t} d^2x \; r_a N_b p^{ab}_\sigma
= \alpha_a P^a - \frac{1}{2} \beta_{ab} M^{ab},
\eeq
where
\beq{30}
P^a = -2 \int_{B_t} d^2x ( r_b p^{ab}_\sigma),
\eeq
\beq{31}
M^{ab} = -2 \int_{B_t} d^2x \;  r_c ( x^a p^{bc}_\sigma - x^b p^{ac}_\sigma).
\eeq
Note that all components of $\alpha$ and $\beta$ are functions of $t$
only. In addition we define
\beq{32}
P^0  =  \frac{1}{8\pi G}\int_{B_t} d^2x \sqrt{\sigma}
   \left( k - k_0 \right)  =   - M,
\eeq
where we have used the result of \cite{HaHo} that the integral over
$k_{reg}= k - k_0$ can be identified with the negative ADM-energy $M$.
Let us for simplicity assume here and in the following $\eta = 0 $ 
at $B_\infty= B_t$, then $P^0 = \int_{B_t} d^2x \; \pi$.
The next step is to parametrize this new form of the surface terms
considering $P^0$, $P^a$ and $M^{ab}$ as new momenta. This leads to the
action
\begin{eqnarray}\label{33}
I[{\cal M}, g] &=& \int dt \int_{\Sigma_t} d^3x \left[p^{ab} \dot{h}_{ab}
-N {\cal H} - N_a {\cal H}^a  \right] \\
& & +  \int dt \left[P^0 \dot{\tau}_0 + P^a \dot{\tau}_a +
M^{ab} \dot{\tau}_{ab} - \alpha_0 {\cal C}^0 - \alpha_a {\cal C}^a - \beta_{ab} {\cal C}^{ab}
\right] \nonumber \\
& & +  \int dt \left[ \pi_\theta  \dot{\theta} - \kappa {\cal C}_J
    + \int_J d^2x \left( \pi_J^m \dot{\theta}_m  - N_m {\cal C}^m_J
        \right) \right] \nonumber
\end{eqnarray}
where we have introduced the constraints
\beq{34}
{\cal C}^0 \equiv  P^0 +  M \approx 0,
\eeq
\beq{35}
{\cal C}^a \equiv P^a +  2 \int_{B_t} d^2x \left[ r_b p^{ab}_\sigma \right]
\approx 0,
\eeq
\beq{36}
{\cal C}^{ab} \equiv -\frac{1}{2} M^{ab} -  \int_{B_t} d^2x \;  r_c\left[ x^a p^{bc}_\sigma
- x^b p^{ac}_\sigma \right]  \approx 0.
\eeq
The action (\ref{33}) with the constraints (\ref{7}), (\ref{8}),
(\ref{25}), (\ref{26}), (\ref{34}), (\ref{35}) and (\ref{36}) presents our
main result of this section. It will be used below to discuss thermodynamics
of black holes as well as to formulate the quantized
theory. To be explicit, we will write down also the Hamiltonian
\begin{eqnarray}\label{36.0}
H &=& \int_{\Sigma_t} \left[ N {\cal H} + N_a {\cal H}^a \right] d^3x
    + \int_J N_m {{\cal C}_J}^m d^2x \\
  & &\;\; +  \; \kappa {\cal C}_J + \alpha_0 {\cal C}^0  + \alpha_a {\cal C}^a
      +\beta_{ab}{\cal C}^{ab} \nonumber
\end{eqnarray}
which can, of course, be immediately read off from (\ref{33}).

\section{Thermodynamics and Boundary States}

In this section we show how thermodynamics can be extracted from
the Hamiltonian (\ref{36.0}). To hold our discussion as transparent
as possible we shall restrict our consideration to the particular case
of Kerr black holes. Therefore we assume $N^{\mu}\equiv(N^0, N^a)$ to
asymptotically approach a linear combination of time translation and
rotation (with angular velocity $\Omega$) at infinity:
\beq{TH1}
N^0 \rightarrow N_\infty=1, \hspace{1cm} N^a \rightarrow \Omega 
        {\epsilon^a}_{bc} \varphi^b x^c.
\eeq
This means $\alpha^0=N_\infty, \alpha^a=0,$ and $\beta_{ab}= - \Omega
\epsilon_{abc} \varphi^c$. The particular form of $\beta_{ab}$ contains
\beq{TH2}
\frac{1}{2} \beta_{ab} M^{ab} =  \Omega \int_{B_t} d^2x \; r_c
\epsilon_{abd} \varphi^{d} \left( x^a p^{bc}_\sigma - x^b p^{ac}_{\sigma}
      \right) = -\Omega J,
\eeq
where
\beq{TH3}
J = \varphi^d L_d = 2 \varphi^d \int_{B_t} d^2x \; r_c \epsilon_{dba} x^a
p^{bc}_\sigma
\eeq
is the total angular momentum \cite{ReTe}.
Thus (\ref{33}) can be reduced to
\begin{eqnarray}\label{TH4}
I^{Kerr}[M,g] & = & \int dt \int_{\Sigma_t} d^3x \left[ p^{ab} \dot{h}_{ab}
                -N{\cal H} - N_a {\cal H}^a \right] \\
           &   & \int dt \left[ P^0 \dot{\tau}_0+ P^\omega \dot{\tau}_\omega
                + \pi_\theta \dot{\theta} - N_\infty {\cal C}_0
                - \Omega {\cal C}_\Omega - \kappa {\cal C}_J \right] \nonumber
\end{eqnarray}
with
\beq{TH5}
{\cal C}_\Omega \equiv -P^\omega + J \approx 0.
\eeq
Let us next comment on the variation of (\ref{TH4}). The variation with
respect to the variables $p^{ab}$ and $h_{ab}$ leads to the
equations of motion and a set of boundary terms which are cancelled by
the variation of $M$, $J$, and $A$. The variation of  the
various Lagrange multipliers produces the consraints ${\cal H}_{inner}$
and ${\cal C}_{outer}$.
The conservation laws $\dot{M} = 0$, $\dot{J} = 0$, and $\dot{A}=0$
follow from $\delta \dot{\tau}_0$, $\delta \dot{\tau}_\omega$,
and $\delta \dot{\theta}$, respectively. Finally, the variation
of the boundary momenta $P^0$, $P^\omega$, and $\pi_\theta$ provides the
relations $\dot{\tau}_0 = N_\infty$, $\dot{\tau}_\omega = -\Omega$, and
$\dot{\theta}= \kappa$. While the meaning of the Lagrange multipliers
$N_\infty$ and $\Omega$ is well-understood, the meaning of $\kappa$ remains
unclear. One is thus left with the question what quantity is related
to the change of the boost parameter $\theta$. An 
answer to this question can be found from the boundary term of the 
bifurcation point and requires a computation of the extrinsic curvature.
To perform this calculation it is convenient to use the relation
\beq{TH6}
\Theta = -\nabla_\alpha r^\alpha,
\eeq
where $r^\alpha$ denotes the outward pointing unit normal to the boundary
of the hypersurface $\Sigma$ at the bifurcation point.
Note that the normal $r^\alpha$ lies in the tangent space of 
$\Sigma$ and is pointing by definition in negative $r$-direction. This 
explains the minus sign on the right hand side of equation (\ref{TH6}).
Since the evaluation of the boundary term at the bifurcation point cannot
depend on the foliation between the initial ($\Sigma_0$) and final ($\Sigma_1$)
hypersurfaces we assume for the following that near the bifurcation
point the time flow vector field is given by the Killing vector field $\chi$,
which is associated with any bifurcation horizon.
The lapse function $N$ and the shift vector $N^a$ thus satisfy
\beq{36j}
\chi = N n + \vec{N},
\eeq
where $n$ is the unit normal vector of the hypersurfaces $\Sigma$.
Since $\chi$ vanishes at the bifurcation surface, (\ref{36j}) implies
\beq{36k}
N|_J=0, \;\;\;\;\; N^a|_J=0.
\eeq
Moreover, it can be shown \cite{Brown}
\beq{36l}
\left. \frac{N^a}{(-\chi^\alpha \chi_\alpha)}\right|_J=0 , \;\;\;\;\;
\left. \frac{N^a}{N} \right|_J=0
\eeq
and
\beq{36m}
\nabla_\alpha N|_J = -\kappa r_\alpha,
\eeq
where $\kappa$ denotes the surface gravity.
Thus, when we approach the bifurcation point from an arbitrary hypersurface
$\Sigma$, the shift vector vanishes more rapidly than the lapse function.
Therefore, $\chi$ becomes orthogonal to $\Sigma$ at the bifurcation surface,
so that
\beq{36n}
\chi^\alpha r_\alpha|_J=0.
\eeq
As $(-\chi^\alpha \chi_\alpha)^{1/2} = \sqrt{N^2-N^a N_a}$, the gradient of
$(-\chi^\alpha \chi_\alpha)^{1/2}$ is given by
\beq{36o}
\nabla_\alpha (-\chi^\beta \chi_\beta)^{1/2} =
  \frac{N \nabla_\alpha N}{(-\chi^\alpha \chi_\alpha)^{1/2}}
  - \frac{N^b \nabla_\alpha N_b}{(-\chi^\alpha \chi_\alpha)^{1/2}},
\eeq
where the second term vanishes at the horizon because of (\ref{36l}).
On the horizon one  has in addition the relation
$
\nabla_\alpha (\chi^\beta \chi_\beta) = -2\kappa\chi_\alpha$
between $\chi$
and the surface gravity $\kappa$ \cite{Wald}.
Using (\ref{36l}) as well as $(-\chi^\alpha \chi_\alpha)|_J=N^2$,
and taking into account
equation (\ref{36m}), one can write for (\ref{36o})
\beq{36q}
\frac{1}{N} \chi_\alpha = - r_\alpha.
\eeq
Note that the last equation only holds on the horizon.
Thus we cannot differentiate this equation by applying $\nabla^\alpha$.
But we can
apply $\chi^{[\gamma} \nabla^{\alpha ]}$ to any equation holding on the
horizon \cite{WaldRe}. Differentiating (\ref{36q}) with $\chi^{[\gamma}
\nabla^{\alpha ]} $ and contracting afterwards with $\chi_\gamma$ leads
to
\beq{36q1}
- \nabla^\alpha r_\alpha =  \frac{\kappa}{N},
\eeq
where also use of the Killing property $\nabla_{(\alpha}\chi_{\beta)}=0$  and
the relation $r^\alpha \nabla^\gamma  r_\alpha=0$ was made.
As expected, the trace of the extrinsic curvature is divergent at the horizon.
Nevertheless, inserting (\ref{36q1}) into the boundary term one gets a
finite result,
\beq{36r}
\int_{S^2 \times I} d^3x \; \Theta \sqrt{-\gamma} = \int_{S^2 \times I} d^3x \;
\kappa \frac{1}{N} \sqrt{-\gamma} = \int dt \; \kappa A(r_+),
\eeq
where the last equality follows from the line element (\ref{3}) by use
of (\ref{36l}).
Comparison of the expressions (\ref{22}) and (\ref{36r}) thus yields the
relation
\beq{36s}
\dot{\theta} = \kappa,
\eeq
which we have anticipated above with the used notation and  
which is the answer to our foregoing question. We see that the boost parameter
connecting the unit normals between two
hypersurfaces $\Sigma_{t_1}$ and $\Sigma_{t_2}$ is given by $\theta =
\kappa (t_2 -t_1)$.
Since the surface gravity $\kappa$ is zero for extreme black holes,
also the surface term at the bifurcation point is vanishing in these cases.
Thus, in the extreme cases the boundary term at the bifurcation point
produces no additional degree of freedom. As this additional degree of freedom
is connected with the entropy of black holes \cite{BrKi}, one may suspect 
that extreme black holes have zero entropy as well as zero temperature.
This point will be discussed further in the next section.

The next goal we shall strive for is the derivation of the first law
of black hole mechanics. This goal can be achieved by a slight 
modification of the perturbation procedure described in \cite{SuWa,Wald}.
Let $[h_{ab}, p^{ab}; \tau_0, \tau_\omega, \theta
,P^0, P^\omega, \pi_\theta)$ be any initial asymptotically flat data set 
satisfying the constraints and let $[\delta h_{ab}, \delta p^{ab}; 
\delta \tau_0, \delta \tau_\omega,$ $\delta \theta,\delta P^0,
\delta P^\omega, \delta \pi_\theta)$ be any smooth asymptotically flat 
perturbation satisfying the linearized constraints. If the Hamiltonian
is a linear combination of constraints, it is obvious that for stationary
spacetimes $\delta H$ must vanish. For our Hamiltonian this means
\beq{1TH1}
\frac{\kappa}{8\pi G} \delta \pi_\theta + N_\infty \delta P^0 + \Omega 
\delta P^{\omega}=0.
\eeq
Note that the perturbation of $h_{ab}, \pi^{ab}$ implies a perturbation
of $M$, $A$, and $J$. Use of the constraints transforms (\ref{1TH1})
into the standard relation
\beq{1TH2}
\delta M = \frac{\kappa}{8\pi G} \delta A + \Omega \delta J.
\eeq
For the more general form of the Hamiltonian (\ref{36.0}) the same
consideration leads to 
\beq{36v}
\kappa \delta \pi_\theta + \alpha_0 \delta P^0 + \alpha_a \delta P^a
-\frac{1}{2} \beta_{ab} \delta P^{ab} + \int_J N_m \delta {\pi_J}^m d^2 x = 0.
\eeq
We have thus found a generalization of the standard formula (\ref{1TH2})
(here given for spacetimes which are Poincar\'{e} invariant). For a discussion of the last term in (\ref{36v}) see \cite{Teitelboim}.

The equation (\ref{1TH2}) (or more general (\ref{36v})) provides the theory 
with the so
called first law of black hole mechanics. The identification of
$\kappa /2\pi$ with the temperature and $A/4G$ with the entropy of
black holes is not achieved by these means. This still relies on the quantum field theory scattering calculation of Hawking \cite{Hawking} or the path integral formulation given by Gibbons and Hawking \cite{GiHa}.

\section{Entropy, Nernst Theorem and Topology}

In this section we shall reformulate the treatment of \cite{GiHa} to define
the entropy of black holes from the above discussed boundary states. 
This will also explain how the vanishing of the boundary term is related 
to the zero entropy interpretation. 
The central point of the following investigation is 
the action evaluated for stationary spacetimes 
($\dot{h}_{ab}=0, \dot{p}^{ab}=0$), which in a gravitating system
must be viewed to describe equilibrium states. On-shell the
constraints are fulfilled and thus (\ref{TH4}) reduces to
\beq{p1}
I= \int dt \left[ \frac{\kappa}{8 \pi G} A(r_+) - M + \Omega J \right].
\eeq
The surface gravity $\kappa$ was found to be \cite{Hawking}
\beq{p2}
\kappa = 2\pi \beta^{-1},
\eeq
where $\beta$ denotes the inverse temperature at infinity. For the
definition of entropy one can now refer to the path integral
formulation. 
The connection between the path integral approach and thermodynamics
is obtained from the transition amplitude $< q_f t_f| q_i t_i> = \int
Dq \exp(i I[q]/\hbar)$, which in the Schr\"odinger picture can also be written 
as $< q_f|\exp[-i\hat{H} (t_f - t_i)/\hbar ] | q_i>$, where $H$ denotes the 
Hamiltonian. In the latter case the partition function can be expressed
by putting $q_f=q_i$, summation over 
a complete set of eigenstates, and identification of $t_f-t_i$ with
$-i\hbar \beta =-i\hbar/k_B T$ (i.e. $Z = tr(\exp[- \hat{H}/k_B T ])$).
Applying the same procedure to the path integral expression of the 
transition amplitude leads to
\beq{p2a}
Z = \int Dq \; e^{-I_E/\hbar},
\eeq
where $I_E$ is the Euclidean action resulting from a Wick rotation
$t\rightarrow i\tau$, and where the integral is over all $q$'s  which
are periodic in $\tau$ with periodicity $\hbar\beta$. In the saddle point
approximation the Euclidean action for (\ref{TH4}) can be easily read off
from (\ref{p1})($\hbar = 1$):
\beq{p3}
I^E= \beta \frac{\kappa}{8 \pi G}  A - \beta M + \beta \Omega J .
\eeq
For extremal black holes the same consideration will lead to a
Euclidean action
\beq{p4}
I^{E}_{ext} = - \beta M + \beta \Omega J.
\eeq
Moreover, in the saddle point approximation (because of (\ref{p2a})) 
the Euclidean action $I^E$ is directly related to the
thermodynamical potential of a grand canonical ensemble via $I^E= - \beta
W$ \cite{GiHa}. The total differential $dW$ calculated from (\ref{p3}) 
therefore reads 
\beq{p4a}
dW = - \frac{A}{4G} d T - J d\Omega,
\eeq
where use was made of (\ref{1TH2}) and where the identification 
(\ref{p2}) with $\beta = T^{-1}$ was put in ($k_B=1$).
Thus in the case of nonextreme black holes one finds the Bekenstein-Hawking
formula for the entropy:
\beq{p4b}
S = - \left(\frac{dW}{dT}\right)_\Omega = \frac{A}{4G}.
\eeq
In the extreme case one has to use the total derivative of the potential 
defined by (\ref{p4}) and thus finds
\beq{p4c}
S_{ext}= -\left(\frac{dW_{ext}}{dT}\right)_\Omega = 0.
\eeq
Hence, in the extreme case one has a vanishing entropy, although the 
horizon area itself is not vanishing. This is a remarkable result, which
follows here from a consistent treatment of the Hamiltonian formulation.

It is also interesting to compare our expression for the thermodynamical 
potential $W$ with the one found by Gibbons and Hawking \cite{GiHa}. 
In their approach only one boundary
(located at infinity) with topology $S_1 \times S_2$ is considered.
Evaluation of the integral over the trace of the extrinsic curvature
$K_{reg} = K -K_0$ leads thereby to the potential
\beq{p4d}
W = \frac{1}{2} M.
\eeq
The equivalence of the expression (\ref{p4d}) with the one discussed above
can be easily derived by use of the Smarr formula
\beq{p4e}
\frac{1}{2}M = \frac{\kappa}{ 8\pi G} A + \Omega J.
\eeq
Given (\ref{p4d}), the entropy can be deduced from the 
standard formula
\beq{p5}
S = \left( \beta \frac{\partial }{\partial \beta} -1 \right)_J I^E.
\eeq
Since the potentials defined by (\ref{p3}) and (\ref{p4d}) are equivalent
one must assume $M$ to be $\beta$-independent
($\partial M/\partial \beta=0$) in the extreme cases. This `missing' relation
between $M$ and $\beta$ in the extreme cases appearing here from consistency
considerations is also observed in \cite{HaHo} and is there made responsible
for a vanishing entropy.

Let us next comment on some topological issues discussed in the
Euclidean approach and connect them to the results derived above.
In the Euclidean theory black hole spacetimes have the topology
$\R^2 \times S^{D-2}$. Near the horizon one can take the line
element to be given by \cite{Teitelboim}
\beq{36A}
ds^2 = d\rho^2 + \rho^2 \theta^2_E d\tau^2 + \gamma_{mn}
(dx^m + N^m d\tau)(dx^n +N^n d\tau),
\eeq
where we have chosen a polar system of coordinates for the description
of $\R^2$. Note that $\rho=0$ describes here the position of the
horizon, $\theta_E$ is the proper angle of an arc in $\R^2$, and $\tau$
denotes the Killing time. Since a boost at the bifurcation point corresponds to a rotation in the Euclidean spacetime, a `Wick rotation'
of the boost parameter $\theta$ gives the Euclidean
opening angle $\theta_E$ (explicitly, $\theta_E = i \theta$).
Thus we can conclude,
\beq{36B}
\theta_E = \kappa (\tau_2-\tau_1)= 2\pi \beta^{-1} (\tau_2 - \tau_1).
\eeq
By demanding regularity for the line element (\ref{36A}), we thus find
immediately the period $\beta$ for the $\tau$-coordinate. This nicely displays
the correlation between regularity of the Euclidean line element and black
hole temperature, as it is used in the path integral approach \cite{GiHa}
discussed above.
For extreme black holes $\theta_E$  vanishes identically and therefore
the metric (\ref{36A})  is no longer well-defined. As a way out one can
define the line element in the extreme cases to read
\beq{36C}
ds^2 = d\rho^2 + e^{\rho^2} d\tau^2 + \gamma_{mn}
(dx^m + N^m d\tau)(dx^n +N^n d\tau).
\eeq
As a consequence of the metric (\ref{36C}) the position of the horizon
is now given by $\rho = - \infty$ and belongs no longer to
the manifold itself. Hence, extreme black holes have the topology of an 
annulus and not of a disk (as the nonextreme ones).
Since the topology of a classical theory is a
quality one tries to preserve under quantization one here must conclude,
 as discussed in \cite{Teitelboim, HaHo}, that the entropy of extremal
black holes is zero. The relation (\ref{36B}) derived from the Lorentzian 
theory in the last section now connects the observed topology change directly 
with the vanishing of the surface gravity $\kappa$.
Since the temperature of black holes
is proportional to the surface gravity, the observed topology change is
thus directly related to the vanishing of the temperature.

More recently, another interpretation for the entropy of black holes was
drawn which is also connected to topological issues but this time in the
Lorentzian theory \cite{BrKi,Martinez}. By investigations of the Penrose 
diagram
it is quite obvious that for nonextreme black holes the data within
call from an arbitrary slice $\Sigma $ starting at the bifurcation sphere
allow not to recover the full spacetime. This drastically differs from
the extreme cases, where the maximum information is already contained
in such a hypersurface. Also here the change in the Penrose diagrams is
caused by the surface gravity $\kappa$.

\section{Canonical Quantization}

For the discussion of the canonical quantum theory of gravity we will
also consider interactions with matter fields. For simplicity we shall here
represent these fields  by a minimally coupled scalar 
field.\footnote{For a non-minimally coupling
-- a coupling of the scalar field not only to the metric but also to
the curvature ${\cal R}$ -- we would have to consider additional
boundary terms. This we shall discuss in section 8.} As a
consequence the Hamiltonian and diffeomorphism constraints acquire
additional contributions ${\cal H}_\phi$ and ${\cal H}_\phi^a$,

\beq{37}
{\cal H} \equiv \frac{1}{2{\cal M}} G_{abcd} p^{ab}p^{cd} + {\cal M}V[h_{ab}]
+ {\cal H}_\phi \approx 0,
\eeq
\beq{38}
{\cal H}^a \equiv -2D_b p^{ab} + {\cal H}_\phi^a \approx 0,
\eeq
where we have for convenience introduced the DeWitt metric
\beq{39}
G_{abcd} = \frac{1}{2\sqrt{h}} \left( h_{ac}h_{bd} + h_{ad}h_{bc} - h_{ab}
h_{cd} \right)
\eeq
and made use of the abbreviations ${\cal M}=1/32\pi G $ and
$V[h_{ab}]=-2 \sqrt{h} (R-2\Lambda)$.
The classical theory is completely described by a set
of constraints: the well-known Hamiltonian and diffeomorphismen constraints,
and a couple of extra constraints arising from boundary terms.

Quantization is now performed in the standard formal manner by
replacing all momenta with $\hbar/i$ times (functional) derivatives and
implementing all constraints by acting on wave functionals
$\Psi[h_{ab}(x), \phi(x), \theta_m(x); \tau, \tau_a, \tau_{ab},
\theta)$.
At this point one normally has to rely on a particular factor
ordering. But since later on only the semiclassical behaviour is considered
explicitly, we do not need to fix this ambiguity.
The constraint equations read:
\beq{39a}
{\cal H}_{inner} \Psi = 0,
\eeq
\beq{39b}
{\cal C}_{outer} \Psi = 0.
\eeq
Note that in contrast to the standard discussion of the Wheeler-DeWitt
approach one now has also to take into account the outer constraints 
(\ref{39b}). 

To define thermodynamical properties one would like to generalize
(\ref{36v}) to
\beq{39c}
\kappa \delta \frac{\partial \Psi}{\partial \theta} 
+ \alpha_0 \delta \frac{\partial \Psi}{\partial \tau}
+ \alpha_a \delta \frac{\partial \Psi}{\partial \tau_a}
- \frac{1}{2} \beta_{ab} \delta \frac{\partial \Psi}{\partial \tau_{ab}}
+ \int_J d^2x \; N_m \delta \frac{\delta \Psi}{\delta \theta_m}
= 0.
\eeq
Using the constraints (\ref{39b}), this would lead to the classical
result. But as was discussed in section 4, the validity of 
equation (\ref{36v}) is based on stationary spacetimes (as well as on 
certain classes of perturbation).
In the classical theory stationarity is provided by the existence of
timelike Killing vector fields.
But what this classical property of black hole spacetimes means in
the quantized theory is quite unclear. Therefore we shall discuss
in the following a semiclassical approximation, where one can at least 
approximately rely on such classical features.

\section{Semiclassical Approximation}

Starting point for the semiclassical approximation is the distinction
between `nearly' classical and `fully' quantum theoretical quantities.
This can
be compared with a similar situation in molecular physics, where a
dynamical distinction between the heavy nuclei and the light electron
can be profitably used to develop the Born-Oppenheimer approximation.
For our case we assume gravity to take over the role of the
slowly moving nuclei, while the matter field takes over the role of the
fast moving electrons. As is shown in \cite{Kiefer}, this qualitative
difference can be formally introduced into the approximation scheme
by choosing a WKB-type ansatz for the state functional
\beq{40}
\Psi = \exp \left( \frac{i}{\hbar} S \right)
\eeq
and by expansion of $S$ in powers of ${\cal M}$,
\beq{41}
S = {\cal M} S_0 + S_1 + {\cal M}^{-1} S_2 + ... \; \; .
\eeq
In the highest order (${\cal M}^2$) one finds that $S_0$ is independent
of $\phi$, i.e. independent of the matter
fields. In the next order (${\cal M}^1$)
one gets the Hamilton-Jacobi equation for the gravitational field
\beq{42}
\frac{1}{2} G_{abcd} \frac{\delta S_0}{\delta h_{ab}}
\frac{\delta S_0}{\delta h_{cd}}+ V[h_{ab}] = 0.
\eeq
This equation (together with the corresponding diffeomorphism equation)
is equivalent to all ten of Einstein's field equations (here in vacuum)
and therefore
describes the classical solutions of general relativity --
in our case the classical black hole solutions. The solution of this
equation can also be used to discuss the meaning of classically allowed
and forbidden regions \cite{BrKi}.

In contrast to the standard discussion one finds in addition to the
Hamilton-Jacobi equation (\ref{42}) the constraints:
\beq{43}
\frac{\partial S_0}{\partial \theta} - \frac{A(r_+)}{8\pi G} = 0\;,
\eeq
\beq{44}
\frac{\delta S_0}{\delta \theta_m} +2 r_n p^{mn}_\sigma = 0\;,
\eeq
\beq{45}
\frac{\partial S_0}{\partial \tau_a} + 2\int_{B_t} d^2x
  [r_b p_\sigma^{ab}] = 0\;,
\eeq
\beq{46}
-\frac{1}{2}\frac{\partial S_0}{\partial \tau_{ab}} - \int_{B_t}
d^2x \; r_c [x^a p_\sigma^{bc}-  x^b p_\sigma^{ac}] = 0\;,
\eeq
\beq{47}
\frac{\partial S_0}{\partial \tau_0} + M = 0\;.
\eeq
They are easily solved by a separation ansatz
\beq{48}
S_0[h_{ab}(x), \theta_m(\tilde{x}); \tau_0 , \tau_a, \tau_{ab}, \theta)
= S_0[h_{ab}]+ S_0[\theta_m] + S_0(\tau_0) + S_0(\tau_a)
+ S_0(\tau_{ab}).
\eeq
As a consequence of the discussion above, the boundary contributions to 
the Hamilton-Jacobi functional
are connected with the thermodynamical properties of black holes.
Using equation (\ref{36v}) one finds
\beq{48a}
\kappa \delta \frac{\partial S_0}{\partial \theta} + \alpha_0 \delta
\frac{\partial S_0}{\partial \tau_0} + \alpha_a \delta
\frac{\partial S_0}{\partial \tau_a} - \frac{1}{2} \beta_{ab} \delta
\frac{\partial S_0}{\partial \tau_{ab}} + \int_J N_m \delta
\frac{\partial S_0}{\partial \theta_m} d^2 x = 0.
\eeq
Note that this relation only holds if the classical black hole spacetime
described by $S_0$ possesses a timelike Killing vector field.
For equilibrium states, such a timelike Killing vector field must exist
for every order of our approximation scheme. Thus, the 
generalization of (\ref{36v}) from order to order should lead to the
corresponding thermodynamical description.

In the next order (${\cal M}^0$) one finds by introducing the wave functional
\beq{49}
\chi = D[h_{ab}] \exp\left( iS_1/\hbar \right),
\eeq
the local functional Schr\"odinger equation
\beq{50}
i\hbar G_{abcd} \frac{\delta S_0}{\delta h_{ab}}
\frac{\delta \chi}{\delta h_{cd}} \equiv i \hbar
\frac{\delta \chi}{\delta {\cal T}} = {\cal H}_\phi \chi
\eeq
for quantum fields propagating on the classical spacetime
described by $S_0$. The functional $D$ is thereby chosen in such
a way that it obeys the standard WKB prefactor equation \cite{Kiefer}.
Taking into account the boundary terms we find that $S_1$ is
independent of all gravitational boundary variables, while surface terms of
the matter fields would lead to an additional contribution.
If there are no additional surface terms
of the quantum fields, the thermodynamical properties of the black
hole will remain unchanged at first glance.

Until now there is one weak point in our discussion -- our fixation
procedure of the hypersurface at the bifurcation point presupposes the
knowledge of its position. This horizon position $r_+$ we had to put in
by hand. This intervention by hand also fixes  the thermodynamical
properties of our theory. One thus has to find a way to define the position
of the horizon in every order of approximation. In the order ${\cal M}^0$
this can be achieved with the semiclassical Hamilton-Jacobi equation 
discussed in \cite{Kiefer}.
To define this `corrected' equation one has to decompose the functional
$\chi$ into
\beq{51}
\chi \equiv C \exp\left(i\vartheta\right)
\eeq
and write the Hamilton-Jacobi equation (\ref{42}) up to order
${\cal M}$ in the form
\beq{52}
\frac{1}{2{\cal M}} G_{abcd} \left( {\cal M} \frac{\delta S_0}{\delta
h_{ab} } + \frac{\delta \vartheta}{\delta h_{ab}} \right)
\left( {\cal M} \frac{\delta S_0}{\delta h_{cd}}
+ \frac{\delta \vartheta}{\delta h_{cd}} \right) + {\cal M} V[h_{ab}]
-\frac{\delta \vartheta}{\delta {\cal T}} + {\cal O}({\cal M}^{-1}) = 0.
\eeq
Next we take the expectation value with respect to $\chi$ and use the
relation
\beq{53}
<\! \chi| \frac{\delta \vartheta}{\delta {\cal T}}\chi \!> = - <\! \chi|
{\cal H}_\phi \chi \! >
\eeq
to write
\beq{54}
\frac{1}{2{\cal M}} G_{abcd} \Pi^{ab} \Pi^{cd} + {\cal M} V[h_{ab}] +
<\! \chi| {\cal H}_\phi \chi\! > + {\cal O}({\cal M}^{-1})= 0,
\eeq
where we have introduced the notation
\beq{55}
\Pi^{ab} \equiv {\cal M} \frac{\delta S_0}{\delta h_{ab}}
+ <\! \chi| \frac{\delta \vartheta}{\delta h_{ab}} \chi \!>.
\eeq
As is argued in \cite{Kiefer}, one has in this order of approximation to
interpret the momenta $\Pi^{ab}$ as the geometrodynamical momenta.
Therefore the position of the horizon has now to be derived from the 
`back reaction corrected' Hamilton-Jacobi functional 
$\widetilde{S}_0$ given by
\beq{56}
\frac{1}{2{\cal M}} G_{abcd} \frac{\delta \widetilde{S}_0}{\delta h_{ab}}
\frac{\delta \widetilde{S}_0}{\delta h_{cd}} + {\cal M} V[h_{ab}] +
<\! \chi| {\cal H}_\phi \chi\! > + {\cal O}({\cal M}^{-1}) = 0.
\eeq

Note that in the above investigation we assumed the wave functional to be in a
particular WKB-state. But since the fundamental equations are linear,
one would expect arbitrary superpositions of such states to occur. 
It can easily be seen that such superpositions are only possible for
`inner' states (solutions of the inner constraints). The outer constraints
do not allow for superpositions and must be interpreted as describing 
superselection rules (fixing mass, angular momentum, etc.).
Nevertheless, since the wave functional at this order of approximation reads
\beq{56a}
\Psi \approx e^{iS_0}\chi,
\eeq
one could, for example, also discuss the superposition
\beq{56b}
\Psi \approx \left( e^{iS_0[h_{ab}]} \chi + e^{-iS_0[h_{ab}]} 
     \bar{\chi}\right) e^{i S_0(\theta,\tau_a,...)}.
\eeq
This state could naivly be called a ``superposition of a black hole with a 
white hole''.
As was shown in \cite{DeKi} for a two-dimensional dilaton gravity model,
such superpositions are suppressed by decoherence.
Decoherence is caused by the presence of a huge number of irrelevant degrees of freedom. In \cite{DeKi} it is argued
that even the correlation between Hawking radiation (viewed there as the 
irrelevant degrees of freedom) and the black hole 
can lead to decoherence. Thus in this sense the various semiclassical 
components should become dynamically independent by radiation of the
black hole itself. This would give additional support to the above
consideration of particular WKB-states.

After this very general discussion one would now like to consider
the particular example of 3+1 dimensional black hole solutions.
But since no consistent derivation of back reaction exists in four
dimensions, one has to look out for a more simple model.
As we will see in the next section such a model exists in 2+1 dimensional
gravity.

\section{BTZ Black Holes as an Example}

\subsection{Classical Description}

Some years ago a black hole solution in $2+1$-dimensional  gravity was 
discovered \cite{BTZ}. Since these BTZ black holes are embedded in a 
`well-understood' 3-dimensional gravity theory, they have provoked 
particular interest in the last few years \cite{Carlip}. Recently attempts 
were 
made to study back reaction on their geometry by matter fields \cite{Steif}.
An exact solution of the corresponding semiclassical Einstein equation 
was found for the special case of conformal fields \cite{MaZa}. 
For this reason, $2+1$ gravity provides a good 
model for our purposes. To connect this model 
with our discussion above, we shall first of all construct the corresponding 
Hamilton-Jacobi functional for the classical, matter free theory. 

We start from the action
\beq{BTZ1}
I= \frac{1}{2} \int d^3x \sqrt{-g} \left[\frac{{\cal R} + 2l^{-2}}{\lambda}
   \right] + I_b
\eeq
without any matter field. Note, that $-l^{-2}$ denotes the
cosmological constant and that $I_b$ stands for an appropriate boundary
term. Note also that in this subsection  we shall for simplicity use units
in which  $\lambda = 8\pi G = 1$.
Given the action, we have to perform the ADM decomposition next.
Let us therefore denote the induced metric on the two-dimensional
hypersurfaces $\Sigma$ by
\beq{BTZ2}
d \sigma^{2}= L^2 dr^2 + 2 Q^2 dr d\phi +R^2 d\phi^2.
\eeq
Restriction to axisymmetric solutions then leads after some calculations 
to the constraints
\beq{BTZ3}
\tilde{{\cal H}} \equiv -\frac{ \Pi_L \Pi_R}{LR} + \frac{1}{4Q^2} \Pi_Q^2
   - \frac{1}{(\det \sigma)^2} \left[ Q^6 \frac{d}{dr}\left(\frac{RR'}{Q^2}
   \right) - R^3L^3 \frac{d}{dr} \left(\frac{R'}{L} \right) \right] - l^{-2} 
   \approx 0,
\eeq
\beq{BTZ4}
\tilde{{\cal D}}^{r} \equiv \frac{1}{2} Q'\Pi_Q - \frac{1}{2} Q \Pi_Q'
            - L \Pi_L' + R' \Pi_R \approx 0,
\eeq
\beq{BTZ5}
\tilde{{\cal D}}^{\phi} \equiv \left(\frac{Q^2}{L} \Pi_L \right)' 
             + \left( \frac{R^2}{2Q} \Pi_Q \right)' \approx 0.
\eeq
The last constraint can immediately be integrated and gives the
relation
\beq{BTZ6}
\Pi_Q = -\frac{2Q^3}{LR^2} \Pi_L + \frac{Q}{R^2}J,
\eeq
where $J/2$ is the corresponding integration constant.
Inserting this relation into the Hamilton constraint (\ref{BTZ3})
and demanding  $Q\equiv 0$ the above constraint system reduces to
\beq{BTZ7}
{\cal H} \equiv - \Pi_L \Pi_R + \frac{L}{4R^3} J^2
   +\frac{R''}{L} - \frac{L'R'}{L^2} - LR l^{-2}\approx 0,
\eeq
\beq{BTZ8}
{\cal D}^{r} \equiv - L \Pi_L' + R' \Pi_R \approx 0.
\eeq
Note that the restriction $Q\equiv 0$, which leads to a simplification of the 
constraint system, can only be done after inserting (\ref{BTZ6}) into
(\ref{BTZ3}). 
For the new constraints one can construct the corresponding Hamilton-Jacobi
functional in the way described in \cite{BrKi}. It reads
\beq{BTZ9}
S_0=\pm LB \mp R' \ln \left[ 2R'\left(B+\frac{R'}{L} \right) \right],
\eeq
where
\beq{BTZ10}
B =  \left(\frac{R'^2}{L^2} + M - \frac{J^2}{4R^2} - \frac{R^2}{l^2}
     \right)^{1/2},
\eeq
and where the integration constant $M$ can later be identified with 
the ADM-mass. Note that for $\lambda = 1$ the ADM-mass $M$ is dimensionless.
Remarkably, the ADM-mass can on-shell be read off from the functional
\beq{BTZ10a}
M(r) = \Pi_L^2 + \frac{J^2}{4R^2} - \left(\frac{R'}{L}\right)^2 + 
\frac{R^2}{l^2}.
\eeq
This functional is an exact analogon to expressions found for 1+1 
dimensional dilaton gravity and for the spherically symmetric sector
of Einstein gravity \cite{BrKi}. It can be used to determine
the position of the horizon for arbitrary parametrizations. One obtains the
condition \cite{BrKi}
\beq{BTZ11}
\left(\frac{R'}{L}\right)^2 - \Pi_L^2 = 0.
\eeq
Once the Hamilton-Jacobi functional is given, it is straightforward to 
deduce an explicit expression for the metric. Choosing $R(r)=r$ one
recovers the BTZ black hole solution \cite{BTZ}
\beq{BTZ12}
ds^2 = - F dt^2 + F^{-1} dr^2 + r^2 ( N^\phi dt + d\phi)^2
\eeq
with
\beq{BTZ13}
F= - M + \frac{r^2}{l^2} + \frac{J^2}{4r^2},
\eeq
\beq{BTZ14}
N^{\phi} = - \frac{J}{2 r^2}.
\eeq
The black hole horizon following from (\ref{BTZ11}) or (\ref{BTZ13})
reads
\beq{BTZ15}
r_{\pm}= l \left[ \frac{M}{2} \left( 1 \pm \left[ 1 - \left( \frac{J}{Ml} 
         \right)^2 \right]^{1/2} \right) \right]^{1/2}.
\eeq
The entropy is determined by the boundary term which here reads 
\beq{BTZ15a}
I_b= -2\pi \int dt \;\left[\pi_\theta \dot{\theta} + P^0 \dot{\tau}_0 
+ P^\omega \dot{\tau}_\omega - \kappa {\cal C}_\theta - N {\cal C}_0 -N^\phi {\cal C}_\omega \right]
\eeq
with
\beq{BTZ15b}
{\cal C}_\theta \equiv \pi_\theta - 2r_+ \approx 0,
\eeq
\beq{BTZ15c}
{\cal C}_0 \equiv  P^0 + M \approx 0,
\eeq
\beq{BTZ15d}
{\cal C}_j \equiv -P^\omega + J \approx 0.
\eeq
Integration of $t$ up to $\beta = 2\pi / \kappa$ leads then to
an entropy
\beq{BTZ16}
S = 4\pi^2 r_+.
\eeq
Thus, for $J=0$ the entropy is explicitly given by $S = 4 \pi^2 \sqrt{M} l$.
From the standard relation $T= - (g_{tt}')_{r_+}/4\pi$ $(g_{tt} g_{rr}=-1)$
the temperature is ($J= 0$):
\beq{BTZ16a}
T = \frac{\sqrt{M}}{2\pi l}.
\eeq
Note that the Smarr formula for the BTZ solutions reads $2M = TS$, 
which follows immediately from the Euler theorem for homogeneous
functions.

\subsection{Conformally coupled Matter Fields}

To introduce (conformally coupled) matter fields we consider the following 
action: 
\beq{BTZ17}
I=  \frac{1}{2} \int d^3x \sqrt{-g} \left[ \frac{R+ 2l^{-2}}{\lambda}-
      g^{\mu \nu} \nabla_\mu \psi \nabla_\nu \psi - \frac{R}{8} 
      \psi^2  \right].
\eeq
As in the last section one could start to perform the ADM decomposition.
But this leads to a rather complicated system of constraints, for which 
we weren't able to construct the Hamilton-Jacobi functional. On the other
hand  for a consistent choice of $\psi$ \cite{MaZab} one finds in the
spherically symmetric case the solutions
\beq{BTZ17a}
ds^2 = - F(r) dt^2 + F(r)^{-1} dr^2 + r^2 d\theta^2,
\eeq
with
\beq{BTZ17b}
F(r) = \frac{1}{l^2}\left[ r^2- 3B^2 - \frac{2B^3}{r} \right],
\eeq
\beq{BTZ17c}
\psi(r) = \sqrt{\frac{8B}{\lambda (r+B)}}.
\eeq
$B$ is here an arbitrary constant. The horizon of these solutions is at
\beq{BTZ17d}
r = 2B\equiv r_+,
\eeq
and the thermodynamical quantities are 
\cite{MaZab}
\beq{BTZ17e}
T = \frac{9r_+}{16\pi l^2},\hspace{1cm}
S = \frac{8\pi^2 r_+}{3\lambda}.
\eeq
In \cite{MaZab} it is noted `that the entropy differs by a factor $2/3$
from the ``area law'' $4\pi^2r_+/\lambda$'.
Our above consideration now suggests that this deviation from the ``area law''
is a consequence of the conformal coupling of the matter field.\footnote{
The same conclusion follows from the Noether charge consideration of 
black hole entropy \cite{Wald, WaldNO, WaIv}} In this
case one has to take into account  an additional boundary term, which leads to
\beq{BTZ17g}
S = \frac{4\pi^2 r_+}{\lambda} \left( 1 - \frac{\psi(r_+)^2}{8} \lambda 
\right) =  \frac{8\pi^2 r_+}{3\lambda}.
\eeq
Note that in our notation of the last section we have $\lambda = 1$, whereas in
\cite{MaZab} one has $\lambda = 1/8\pi$. Note also that again the Smarr formula
$2M = TS$ holds.

\subsection{Semiclassical Description and Back Reaction}

To introduce back reaction we study the matter action (\ref{BTZ17})
with $\lambda = 1$.
The corresponding renormalized expectation value of the stress-tensor
is calculated in \cite{MaZa} and reads for the so-called transparent
boundary condition (in the case $J=0$)
\beq{BTZ18}
< {T_\mu}^\nu > = \frac{l_p M^{3/2} {\cal A}(M)}{R^3} diag(1,1,-2),
\eeq
with $l_p=\lambda/8\pi$ and where
\beq{BTZ19}
{\cal A}(M) \equiv \frac{1}{2\sqrt{2}} \sum_{n=1}^{\infty} 
     \frac{\cosh 2n \pi \sqrt{M} + 3}{(\cosh 2n\pi \sqrt{M} -1 )^{3/2}}.
\eeq
This leads to a contribution to the Hamilton constraint (\ref{BTZ7}).
This additional contribution can immediately be incorporated if one 
recalls that the Hamilton constraint can be read off from the 
projection of the Einstein equation onto the normal to $\Sigma$. One gets
\beq{BTZ20}
\bar{{\cal H}} \equiv - \bar{\Pi}_L \bar{\Pi}_R 
   +\frac{R''}{L} - \frac{L'R'}{L^2} - LR l^{-2} - \frac{L}{R^2} l_p 
     M^{3/2}{\cal A}(M) \approx 0,
\eeq
\beq{BTZ21}
\bar{{\cal D}}^{r} \equiv - L \bar{\Pi}_L' + R' \bar{\Pi}_R \approx 0,
\eeq
where the use of `bared' quantities is meant to distinguish between the 
`uncorrected'
(\ref{BTZ7},\ref{BTZ8}) and the back reaction-corrected treatment.
The corresponding Hamilton-Jacobi solution now reads
\beq{BTZ22}
\bar{S}_0=\pm L\bar{B} \mp R' \ln \left[ 2R'\left(\bar{B}+\frac{R'}{L} \right)
   \right]
\eeq
where
\beq{BTZ23}
\bar{B} =  \left(\frac{R'^2}{L^2} + M  - \frac{R^2}{l^2}
                    + \frac{2l_p}{R} M^{3/2} {\cal A}(M) \right)^{1/2}.
\eeq
Interestingly enough the value of the ADM mass is not changed by
the back reaction of the matter field \cite{MaZa}.
{}From (\ref{BTZ23}) we find for the horizon 
\beq{BTZ24}
\bar{r}_+ = \sqrt{M} l \left( 1 + \frac{l_p}{l} {\cal A}(M) \right),
\eeq
where the second term is a $l_p$ correction in the sense of the 
above discussed semiclassical approximation scheme. 
Thus the entropy reads
\beq{BTZ25}
\bar{S}= 4\pi^2 \bar{r}_+
         \left(1- \frac{<\psi^2>}{8} \right).
\eeq
Note that we have introduced the boundary term of the conformal coupling 
by replacing $\psi^2$ with  $<\psi^2>$, since otherwise this term
would lead to an inconsistency with the quantum theoretical description
of the matter field\footnote{The same ansatz was suggested recently 
from a different point of view \cite{Frolov}.}. 
The temperature for the black hole solution 
(\ref{BTZ22}) is, up to order $O(l_p^2)$, 
\beq{BTZ26}
\bar{T}=\frac{\sqrt{M}}{2\pi l} \left(1 + 2\frac{l_p}{l} {\cal A}(M)\right).
\eeq
Since ${\cal A}(M)$ is a complicated expression in $M$, the
entropy $\bar{S}$ is no longer a homogeneous function  in $M$.
Thus one cannot expect the Smarr formula $2M=TS$ to hold anymore.
On the other hand, the first thermodynamical relation 
$dM = \bar{T} d\bar{S}$ must be valid by construction.
This leads to the consistency condition $d\bar{S}/dM = 1/T$, which
determines the value of $<\psi^2>$ at the horizon. 
Evaluation of $\tilde{S} = \int dM T^{-1}$ up to order ${\cal O}(l_p^2)$
and comparison with the expression (\ref{BTZ25}) for $\bar{S}$ in the
case of $M \gg 1$, i.e. ${\cal A}(M) \approx \exp(-\pi \sqrt{M})/2$, gives
\beq{BTZ26a}
<\psi^2>_{r_+} \approx 4 \frac{l_p}{l} \exp(-\pi \sqrt{M})\left(1-\frac{2}{\pi \sqrt{M} l}\right).
\eeq
Thus the entropy reads
\beq{BTZ26aa}
\bar{S} = 4\pi^2 \sqrt{M} l + 4\pi \frac{l_p}{l} 
\exp\left(-\pi\sqrt{M}\right) + {\cal O}(l_p^2)
\eeq
Recall that `per definition' our semiclassical description is only valid if 
gravity behaves classically. This naturally implies the considered limit 
$M \gg 1$.

Using on the other hand the relation between the two point function
and the appropriate Green function \cite{Steif,LiOr}, 
$<\psi^2> = \frac{1}{2} \lim_{x \rightarrow x'}G_{reg}(x,x')$, one receives
for the considered transparent boundary condition
\beq{BTZ26b}
<\psi^2> =\frac{1}{4\sqrt{2}\pi l} \frac{r_+}{r} \sum_{n \neq 0} 
                  \left\{ \cosh (2\pi \sqrt{M} n) -1 \right\}^{-1/2}
\eeq
For $M\gg 1$ only the $n=\pm 1$ contributions of the sum have to be 
considered. One gets
\beq{BTZ26c}
<\psi^2>_{r_+} \approx \frac{1}{2\pi} \frac{1}{l} \exp(-\pi \sqrt{M})
               = 4 \frac{l_p}{l} \exp(-\pi \sqrt{M}),
\eeq
where we have used $l_p=\lambda/8\pi=1/8\pi$. In the case of $M\gg 1$ this
expression equals (\ref{BTZ26a}). Thus consistency of the 
formula (\ref{BTZ25}) for $M\gg 1$ is shown.

Let us emphasize that our result for the entropy differs from the 
result found in \cite{MaZa}, where no contribution from the non-minimally 
coupled matter field was respected. But this contribution has to be 
taken into account since otherwise the relation $dM=TdS$ does not hold. 

\section{Final Remarks}

In this paper we have discussed the canonical quantization of spacetimes
with boundaries. A key role in our investigation is played by additional
degrees of freedom arising from surface terms. We have shown how these
additional degrees of freedom are connected with standard relations of
black hole thermodynamics. Moreover, we have given a derivation of the
Nernst theorem. In the quantum theory the thermodynamical properties
are rediscovered in the first order of a WKB-like approximation scheme.
Going beyond the first order of this approximation leads to back reaction
effects. We have shown how these back reaction effects alter the 
thermodynamical properties. 
We in particular have discussed how our treatment 
works in the case of the BTZ solution in 2+1 dimensional gravity. Moreover,
this example has provided useful insights into the correlation between 
boundary terms and entropy.

Let us note, that another approach which is in some aspect similar in the 
interpretation of entropy is the Noether charge derivation \cite{Wald,
 WaldNO, WaIv}. From the Noether charge approach one would, for example
deduce similar conclusions we have drawn for the particular
example of the BTZ model. Since the Noether charge interpretation of black
hole thermodynamics is valid for all diffeomorphism invariant theories,
we expect the same for our consideration, although no generalization 
of the WKB approximation for arbitrary diffeomorphism 
invariant theories exists. Whether such a generalization also leads to an 
analog of the semiclassical Hamilton-Jacobi equation is an open issue.
If this is the case one could, of course, capture back reaction in the 
above described way.

Note also that for the discussion of back reaction from Hawking radiation
one has to  generalize the given thermodynamical description to at
least quasi stationary spacetimes. But even in the case of radiating
black holes, the stationarity assumption should be valid as long as
the black hole mass is much bigger than the Planck mass.
\smallskip

As it was recently pronounced \cite{Horowitz} it is somehow ironic
that of all things, new developments in string theory (which seem to 
provide a statistical explanation for black hole thermodynamics) have 
led to a contrary result for the entropy of extremal black holes.
In these developments the statistical interpretation was achieved 
by counting so called BPS states in the weakly coupled string theory.
This counting was 
explicitly done for extremal and nearly extremal black holes and
led to the Bekenstein-Hawking formula $S=A/4G$, whereas for
non extremal black holes a consistent counting is still elusive.
Until now a satisfying explanation for the differing conclusions
between the canonical theory of gravity and string theory has not
been given. Nevertheless, first attempts where made and they seem
to suggest that modification from string scales may play an important role
even if the curvature of the Euclidean solution is everywhere small
(in this case string corrections should be negligible)
\cite{Horowitz}. It would be interesting if such string modifications could
be connected with observations made by Davies \cite{Davies} who found
a phase transition for Kerr black holes at $J/M \simeq 0.68$ and 
Reissner-Nordstr\"om black holes at $q/M \simeq 0.86$. If such a correlation
could be derived it would be a beautiful improvement for the acting in unison
of string theory and general relativity.
Of course, at the same time one again would be faced by the Nernst
theorem, which then would be manifestly broken in string theory.
But since new investigations seems to suggest that the Nernst theorem
is also violated for a ideal boson gas which is confined on a circular
string \cite{Waldxy}, this may indicate that the Nernst formulation
of the third law of thermodynamics will lose its fundamental role in 
string theory.

However, since black hole thermodynamics seems to be the only vague glimpse
we get from the quantum theory of gravity, contrary results between different
approaches are of inestimable value and should be viewed as a chance to
stake off the appropriate camp for quantum gravity.
\bigskip
\bigskip

{\Large \bf Acknowledgement}
\bigskip

It is a pleasure to thank Domenico Giulini and Claus Kiefer for very 
helpful and stimulating discussions. I would especially like to thank 
Claus Kiefer for critically reading the manuscript and making valuable
comments.
Financial supports from the {\it Graduiertenkolleg Nichtlineare 
Differentialgleichungen der Albert-Ludwigs-Universit\"at Freiburg}
are acknowleged.


\end{document}